# Control of high-harmonic generation by tuning the electronic structure and carrier injection

*Hiroyuki Nishidome,[1] Kohei Nagai,[2] Kento Uchida,[2] Yota Ichinose,[1] Yohei Yomogida,[1] Koichiro Tanaka,[2,3] Kazuhiro Yanagi[1*].*

[1]Department of Physics, Tokyo Metropolitan University, Hachioji, Tokyo 192-0397, Japan

[2]Department of Physics, Kyoto University, Sakyo-ku, Kyoto 606-8502, Japan

[3]Institute for Integrated Cell-Material Science (WPI-iCeMs), Kyoto University, Sakyo-ku, Kyoto 606-8501, Japan

ABSTRACT

High-harmonic generation (HHG), which is generation of multiple optical harmonic light, is an unconventional nonlinear optical phenomenon beyond perturbation regime. HHG, which was initially observed in gaseous media, has recently been demonstrated in solid state materials. How to control the extreme nonlinear optical phenomena is a challenging subject. Compared to atomic gases, solid state materials have advantages in controlling electronic structures and carrier injection. Here, we demonstrate control of HHG by tuning electronic structure and carrier injection using single-walled carbon nanotubes (SWCNTs). We reveal systematic changes in the high-harmonic spectra of SWCNTs with a series of electronic structures from a metal to a semiconductor. We demonstrate enhancement or reduction of harmonic generation by more than one order of magnitude by tuning electron and hole injection into the semiconductor SWCNTs through electrolyte gating. These results open a way to control HHG within the concept of field effect transistor devices.

High-harmonic generation (HHG), which is the generation of multiple optical harmonic light, is an unconventional nonlinear optical phenomenon beyond the perturbation regime.[1] HHG has been intensively investigated in gaseous media for application in the generation of attosecond laser pulses, coherent generation of extreme ultraviolet light and soft X-rays, and imaging of molecular orbitals.[2] HHG has also been recently reported in various crystalline solids,[3-5] and HHG in crystalline solids can be used to investigate crystal orientations,[6] quasi-particle dynamics,[7] the Berry curvature,[8] and so on[9]. The generation mechanisms in crystalline solids are different from the HHG in gaseous media because of the periodic arrangements of atoms and collective properties of electrons in solid state materials.[10-12]

There are two distinct dynamics generating high harmonics in solid state materials: interband and intraband dynamics.[5, 10, 13] In the former process, an electron–hole pair is generated by an intense mid-infrared (MIR) field, is simultaneously accelerated by the laser field and subsequently coherently recombines, emitting a high harmonic photon (Figure 1(a)). In the latter processes, electron or hole carriers are driven by the laser field and oscillate in the Brillouin zone, known as dynamical Bloch oscillations (Figure 1(a)).[5, 14] An anharmonic electronic current in the bands results in high-harmonic emission. In both processes, nonlinearity in transport and optical transitions is key for high-harmonic generation, and thus, the electronic structure and carrier density strongly influence both dynamics. In conventional solid state materials, both dynamics will contribute to the high-harmonic generation more or less at the same time. Systematic changes of the bandgap and tuning of the Fermi-level will enable us to understand the underlining physics and find a way to control the HHG in crystalline solids. For this purpose, in this study, we investigated the HHG of single-wall carbon nanotubes (SWCNTs) with a series of different electronic structures and SWCNTs with gate-tuned Fermi-level (Figure 1(b)).

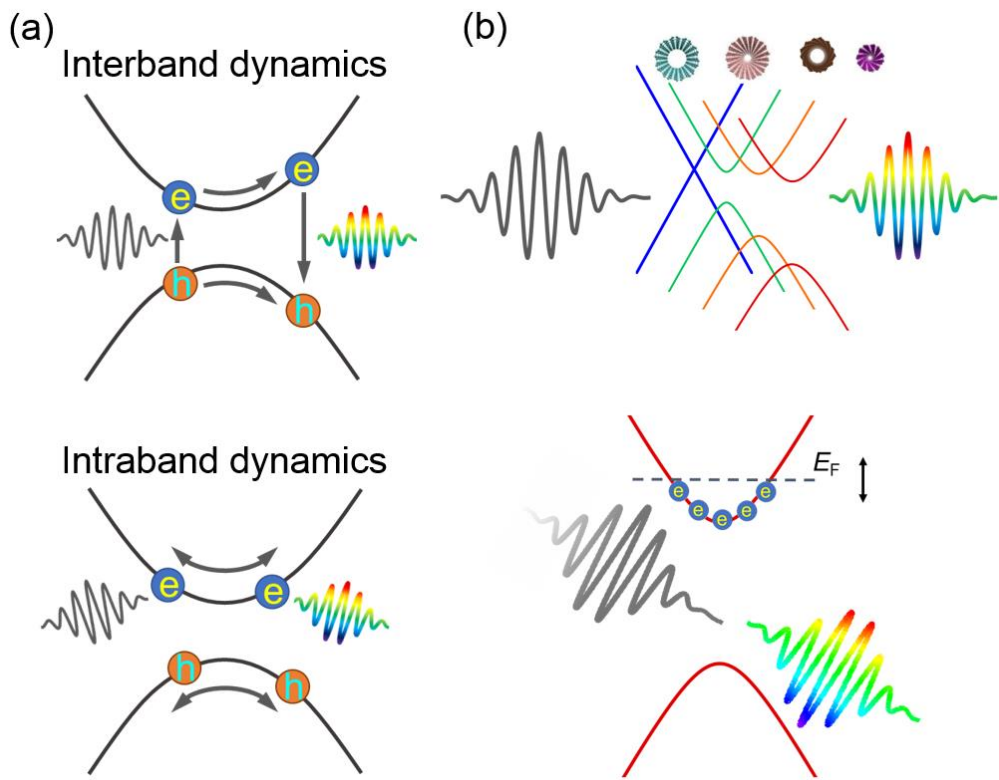

**Figure 1** Interband and intraband dynamics for high-harmonic generation. (a) Schematics of interband dynamics (top) and intraband dynamics (bottom). (b) Control of high-harmonic generation by preparation of SWCNTs with a series of bandgaps (top) and by tuning the Fermi level (bottom).

SWCNTs are rolled graphene tubes, and they show various electronic structures depending on how they are rolled (called chirality). [15] In contrast to related two-dimensional materials, without changing the atomic elements, we can systematically change the bandgaps of SWCNTs from metallic to semiconducting by changing the chirality (Figure 1(b)). In addition, we can tune the Fermi level of SWCNTs by electrolyte gating approaches, where the amount of electron or holes injected through electric double layer formation, is controlled by the shift of the gate voltage (Figure 1(b)).[16-18] In this study, we first investigated how the electronic structures of SWCNTs influence HHG. Then, we investigated how the carrier injection influences HHG spectra using electrolyte gating approaches. The Femi level of SWCNTs can be tuned through carrier injection by electric double layer formation using electrolyte gating. We found that we can control HHG by these approaches.

SWCNTs with different electronic structures were prepared as follows. Metallic SWCNTs (Metal, bandgap energy ($E_g$): 0 eV) and semiconducting SWCNTs with a diameter of 1.4 nm (Semi. #1, $E_g$: 0.7 eV) were prepared by the density-gradient sorting method from SWCNTs produced by the arc discharge method (Arc SO, Meijyo Nano Carbon).[19-21] Semiconducting SWCNTs with an average diameter of 1.0 nm (Semi. #2, $E_g$: 1.0 eV) and (6,5) SWCNTs ((6,5), $E_g$: 1.26 eV) were prepared by gel chromatography.[22] Here, $E_g$ is approximately the optical transition energy between the 1st van Hove singularities of the valence and conduction bands in the semiconducting SWCNTs. The optical absorption spectra of the samples are shown in Figure S1, and we used thin films of the samples. The details of the experimental setup for HHG measurements are described in the Method section. Briefly, we used MIR pulses (central energy: 0.26 eV, pulse width: approximately 60 fs, repetition rate: 1 kHz, maximum pulse energy: ~1 µJ/pulse) generated through different frequency mixing. The MIR pulses were focused onto the sample, and the spot size was estimated to be 60 µm (full width at half maximum of intensity, FWHM). HHG emissions were detected in transmission geometry. For the detection of the 3rd harmonics, we used an InGaAs line detector, and the higher harmonics (>3rd) were detected using a Si CCD camera.

Figure 2(a) shows the relationships between the electronic structure of SWCNTs and the HHG spectra. As shown in the figure, higher order HHG was observed in SWCNTs with larger $E_g$. In the semiconductor SWCNTs with $E_g$ values of 0.7 eV (Semi. #1) and 1.0 eV (Semi. #2), we observed up to the 7th-order harmonics. Moreover, in the SWCNTs with the larger $E_g$ of 1.26 eV ((6,5)), the 11th-order harmonics were clearly observed. In the Metal SWCNTs, we only observed the 5th-order harmonics. In addition, the intensity of the 5th-order harmonics in the Metal SWCNTs was weak compared to that in the (6,5) SWCNTs (see Figure S2 in the S.I.). These results show that the highest order harmonics increases as the bandgap energy increases in the case of SWCNTs.

It is well established that the cut-off energy in HHG is scaled by the peak field strength and photon energy of the pulse laser in solids. [5] Here we experimentally verified the clear change of the highest order harmonics by systematically changing the bandgap energy of the SWCNTs. The observed tendency of the highest order harmonics vs bandgap energy qualitatively agrees with a theoretical model of HHG from a two band system in a solid, [10] in which the cut-off energy linearly increases as the bandgap increases.[10] In addition, the tendency is also in agreement with the recent theoretical study. [23]

We checked whether the observed HHG in SWCNTs corresponds to nonperturbative high harmonics. Figure 2(b) shows the intensities of the 5th - and 7th -order harmonics as a function of laser intensity. As shown here, the power dependence of the HHG intensity does not follow the perturbative nonlinear optics such that the intensity of the $n$-th harmonic generation, $I_n$, is proportional to the $n$-th order of the MIR laser intensity, $I_L^n$. These results indicate the nonperturbative characteristics of the observed HHG in SWCNTs.

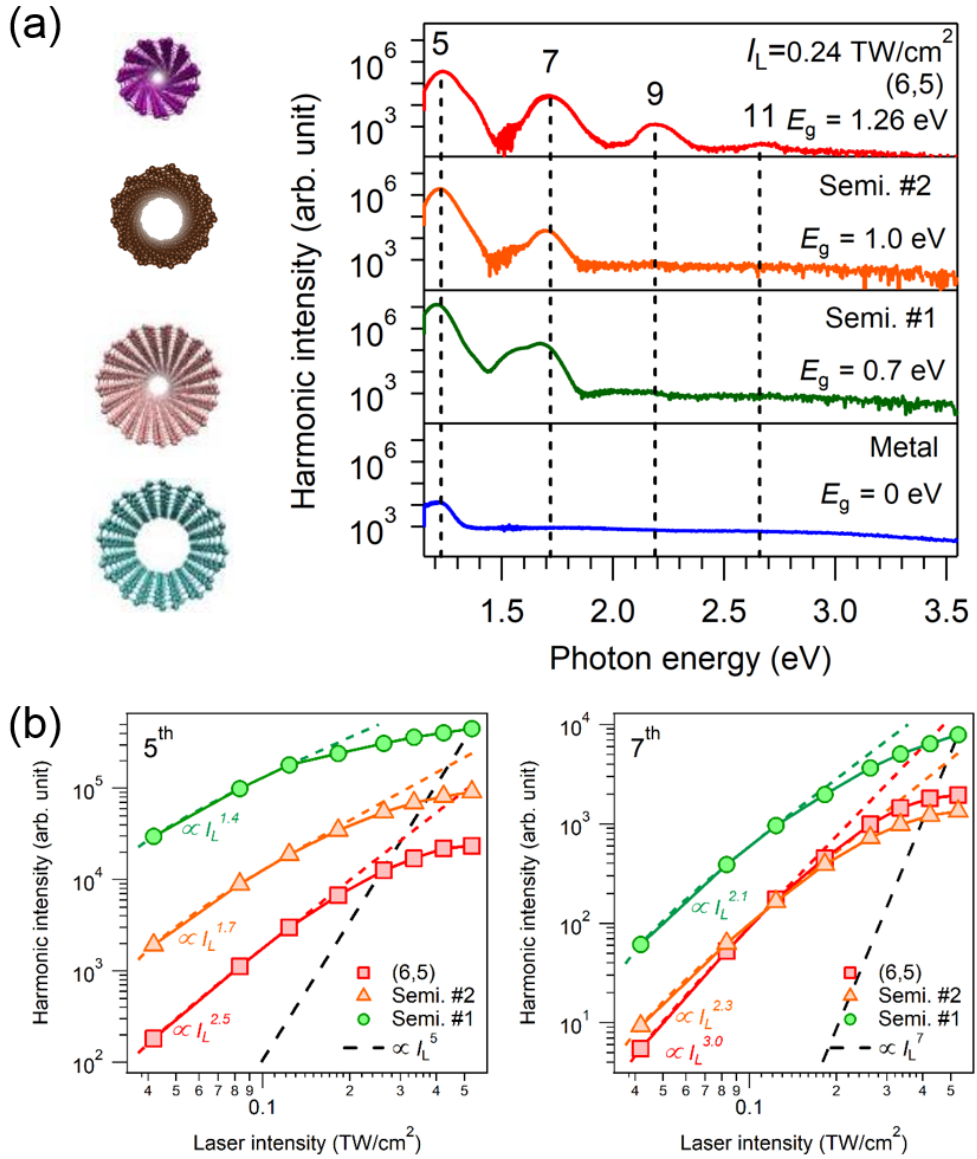

**Figure 2** High-harmonic generation from SWCNTs with a series of electronic structures. (a) High-harmonic generation spectra of (6,5), semiconducting SWCNTs with a diameter of 1.0 nm (Semi. #2), semiconducting SWCNTs with a diameter of 1.4 nm (Semi. #1), and metallic SWCNTs (Metal). All measurements were performed at the laser intensity, $I_L$, of 0.24 TW/cm$^2$. (b) Harmonic intensity of 5$^{th}$- and 7$^{th}$-order harmonics as a function of laser intensity.

When the interband dynamics dominates the HHG in SWCNTs, we expect that band filling by carrier injection will prohibit HHG through Pauli blocking processes, as discussed in photocarrier doping experiments. [24] However, at the same time, we can expect that carrier injection may enhance HHG because of the increase of carriers which can contribute to the generation of high-harmonics. Thus, the influence of carrier injection on the HHG from SWCNTs will not be simple. To experimentally clarify the influence of the carrier on HHG, we investigated HHG in (6,5) SWCNTs, which shows the highest order harmonics in the investigation of bandgap influence, with systematically tuned carrier injections using electrolyte gating techniques. The experimental setup is shown in Figure 3(a). We used an ionic liquid (TMPA-TFSI, Kanto Kagaku Co.), and injected electrons and holes through electric double layer formation by side gating (shift of gate voltage, $V_G$). The details are written in the Method section.

Figures 3(b) and (c) present the HHG spectra from the $3^{rd}$ to $9^{th}$ orders as a function of the positive shift and negative shift of the gate voltage, respectively. Here, as a guide for the location of the bandgap in (6,5) SWCNTs, the optical transition energy between the $1^{st}$ van Hove singularities (1.26 eV) is depicted as dashed lines. A positive shift of $V_G$ injects electrons and a negative shift injects holes into the (6,5) SWCNTs. We adjusted the laser pulse power not to destroy the sample during gating. We could not use a high enough power to observe the $11^{th}$-order harmonics because we were avoiding sample degradation during gating. The reproducibility of the repeated measurements is shown in Figure S3 of the S.I. At the zero gate voltage, we can observe HHG up to $9^{th}$-order harmonics. With a positive shift of the gate voltage, meaning injection of electrons into the nanotubes, the intensity of the $3^{rd}$-order harmonics remarkably increased, whereas that of the $5^{th}$ - and $7^{th}$ -order harmonics decreased. With a negative shift of the gate voltage, meaning injection of holes into the nanotubes, the same behavior was observed, i.e., an increase of the 3$^{rd}$ and decrease of other higher order harmonics. At $V_G$ = −2.1 V or 1.8 V, the intensity of the 3$^{rd}$-order harmonics became almost 10 times larger than that at zero $V_G$, but the intensities of the 5$^{th}$-, 7$^{th}$- and 9$^{th}$-order harmonics completely disappeared. It is noteworthy that the intensities changed by more than one order of magnitude. The details of the changes in the HHG spectra with the shift of the gate voltage are described in Figure S4 of the S.I. Figures 3(d) and (e) show 2D mappings of the HHG spectra and peak intensities of the 3$^{rd}$-, 5$^{th}$- and 7$^{th}$-order harmonics as a function of the gate voltage. These data clearly indicate the presence of two regions: (1) a flat region where all the order harmonic intensities almost do not change with the shift of $V_G$, −1.2 V < $V_G$ < 0.3 V, and (2) a region where enhancement of the 3$^{rd}$-order harmonics and reductions of the 5$^{th}$- and 7$^{th}$-order occurs with the shift of $V_G$, $V_G$ > 0.3 V or < −1.2 V. The boundaries of the two regions are depicted as dotted lines in Figure 3(e), and here, the gate voltages at the boundary are termed threshold voltages. The background of the presence of the threshold voltages is discussed later.

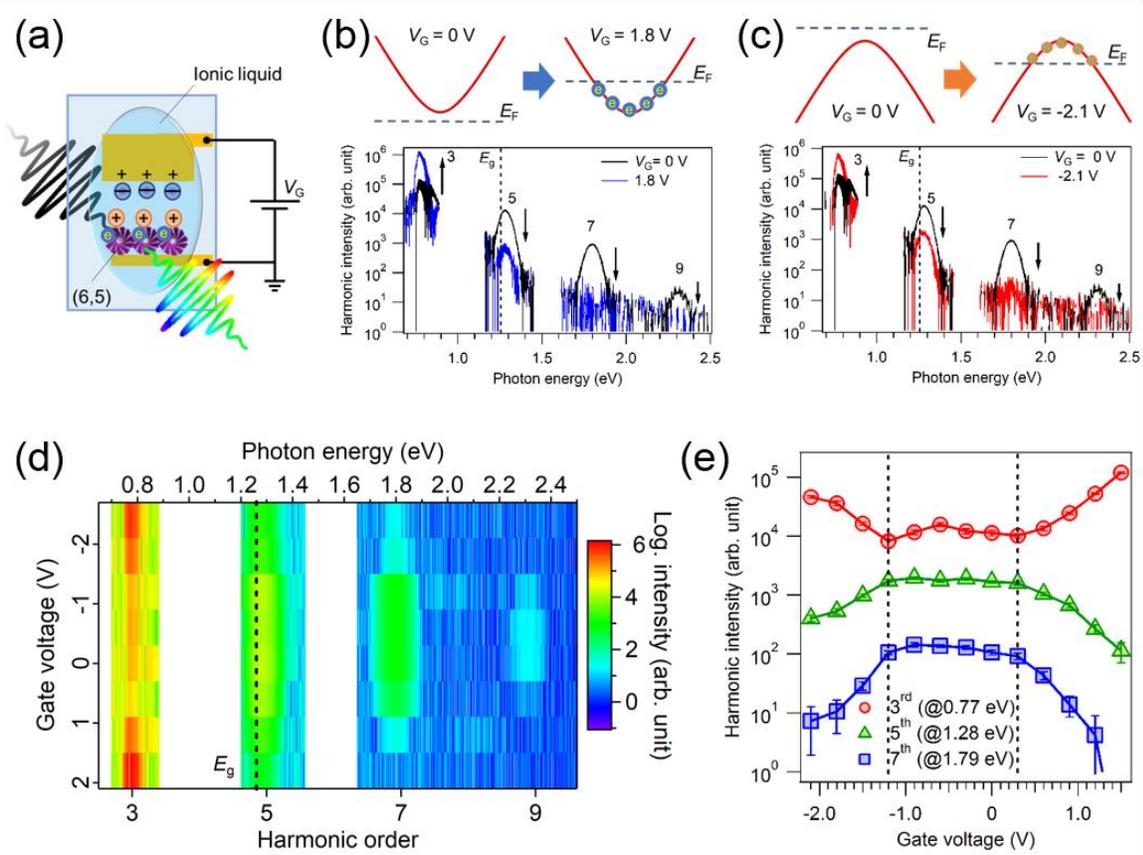

**Figure 3** Gate tuned high-harmonic spectra. (a) Schematic illustration of high-harmonic generation from (6,5) SWCNTs with controlled Fermi level using electrolyte gating. An ionic liquid is used, and carrier injection is controlled by a side gate electrode ($V_G$). (b), (c) High-harmonic spectra with a positive shift, (b) electron injection, and a negative shift, (c) hole injection, of the gate voltage. (d) Intensity mapping of harmonic spectra as a function of the gate voltage. In panels (b), (c) and (d), the location of $E_g$ is plotted as a dotted line. (e) Peak intensities of the 3rd-, 5th- and 7th-order harmonics as a function of the gate voltage. The threshold voltages are marked as dotted lines.

As expected, the decreases in the 5th -, 7th - and 9th -order harmonics suggest a reduction of HHG through the interband dynamics by band filling, i.e. Pauli blocking. [24] In contrast, the increase of the 3rd -order harmonics is caused by an increase of the nonlinear current through the carrier injection. We found that the enhancement occurred in the harmonics whose energy is below the bandgap, and that the reduction occurred in the harmonics whose energy is above the bandgap. To understand the physical mechanism behind these enhancements and reductions in detail, we calculated the HHG spectra using the following model. The details of the model and calculations are given in the Method section and S.I. Here, we assumed an effective two band model of (6,5) SWCNTs, including the 1$^{st}$ conduction and 1$^{st}$ valence bands with a bandgap energy of 1.14 eV within a tight-binding model of hopping energy $\gamma_0 = 3.0$ eV. We calculated the time evolution of both the interband polarization and intraband current contributions in the system during irradiation of a 60 fs MIR laser pulse through a density matrix approach using the Liouville equation. [25] The Fermi-Dirac distribution with chemical potential $\mu$ was assumed as the initial state, and $\mu$ was systematically changed. Here, we set $\mu$ as zero when the $\mu$ was located at the middle of the bandgap. Then, we calculated the harmonic spectra from the Fourier transformation of the real-time dynamics of the two terms. The dephasing time $T_2$ in the calculation influences the calculated intensities of the 3$^{rd}$ -, 5$^{th}$ - and 7$^{th}$ -order harmonics, and we set $T_2$ as 4.4 fs so that the calculated intensities of the 3$^{rd}$ -, 5$^{th}$ - and 7$^{th}$ -order harmonics agree with the experimentally obtained spectrum without the gate voltage (see the details of the calculation method and Figure S5 in the S.I.).

Figure 4(a) shows the calculated 2D mapping of the harmonic spectra, and Figure 4(b) shows the calculated peak intensities of the 3$^{rd}$ -, 5$^{th}$ - and 7$^{th}$ -order harmonics as a function of $\mu$. The calculation qualitatively reproduces the experimental observation: enhancement of the 3$^{rd}$ harmonics and reduction of the 5$^{th}$ and 7$^{th}$ h harmonics when $\mu$ passes across the conduction and valence band edges, which are depicted as dotted lines in Figure 4 (b). Note that the calculation results account for the presence of the experimentally observed threshold voltages for enhancement and reduction of the harmonic intensities. Figures 4(c) and (d) show the intraband current and interband polarization contributions to the $3^{rd}$- and $5^{th}$-order harmonics as a function of $\mu$. When $\mu$ is located within the bandgap, each term of the $3^{rd}$- and $5^{th}$-order harmonics does not change as the shift of $\mu$. However, when $\mu$ approaches the band edge and passes across the edge, in the case of $3^{rd}$-order harmonics, both the intraband current and interband polarization terms increase. This result indicates that injection of carriers enhances nonlinear currents in both terms, and we observed enhancement of the $3^{rd}$ harmonics whose energy is below the bandgap. However, in the case of the $5^{th}$ harmonics, the calculation suggests that the interband polarization term becomes dominant through resonance processes when $\mu$ is located within the band-gap, and the term significantly decreases through Pauli-blocking processes by carrier injection when $\mu$ passes through the band-edge. Although the intraband current term slightly increases by the carrier injection, the reduction of the interband polarization is significant. Therefore, we observed reduction of $5^{th}$ order harmonics. As shown here, our calculation results support experimental observations.

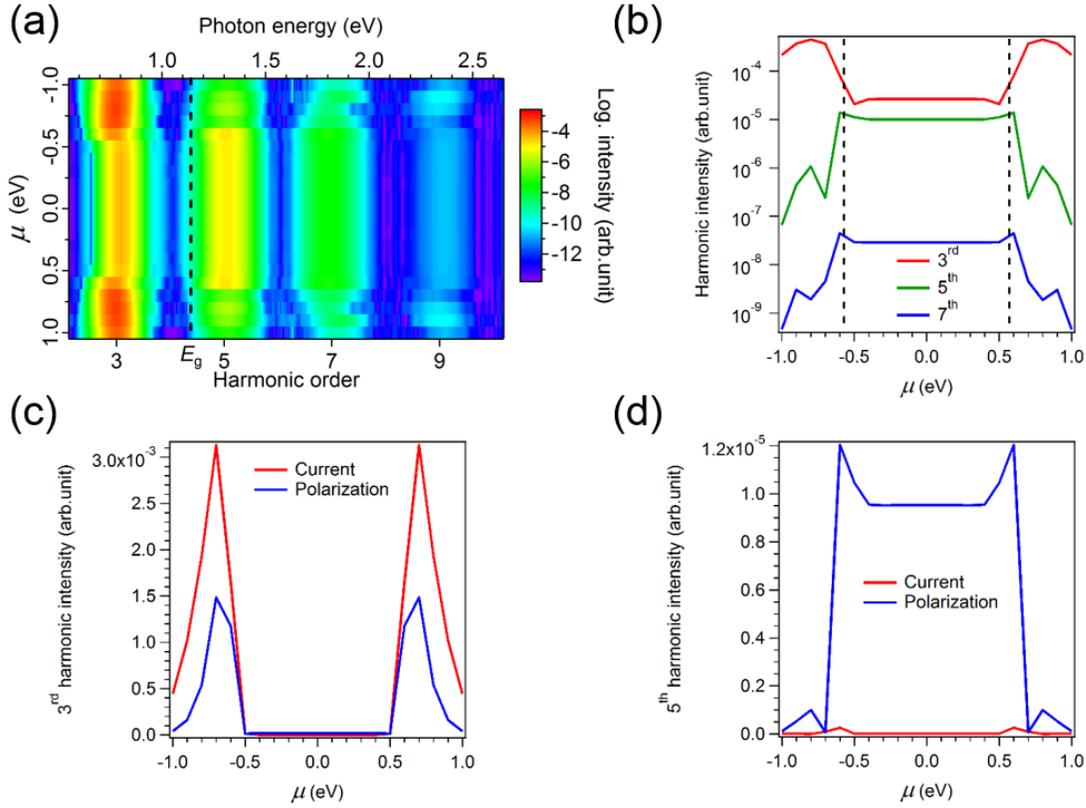

**Figure 4** Calculated high-harmonic spectra. (a) Calculated intensity mapping of harmonic spectra as a function of the chemical potential $\mu$. Here, the bandgap energy, $E_g$, is depicted as a dotted line. (b) Calculated peak intensities of the 3rd-, 5th- and 7th-order harmonics as a function of $\mu$. The locations of the valence band top and conduction band bottom are plotted as dotted lines. (c), (d) Contribution of the current and polarization terms to the (c) 3rd-order harmonics and (d) 5th-order harmonics as a function of $\mu$.

In summary, as demonstrated here, we controlled the high-harmonic generation of SWCNTs by tuning the electronic structure and carrier injections. We can enhance or reduce the high-harmonic intensity depending on the order through the control of carrier injections. Vampa et al. has previously revealed appearance of even order HHG by static electric field applications through symmetry breaking, [26] and our study clarified unique changes of HHG through carrier injections by electric field applications, indicating a way to control HHG by a field effect transistor techniques.

**Method**

**Sample preparation**

Metallic SWCNTs (Metal, bandgap: 0 eV) and semiconducting SWCNTs with a diameter of 1.4 nm (Semi. #1, bandgap: 0.7 eV) were prepared by the density-gradient sorting method from SWCNTs produced by the arc discharge method (Arc SO, Meijyo Nano Carbon Co.). [21] Semiconducting SWCNTs with an average diameter of 1.0 nm (Semi. #2, bandgap: 1.0 eV) and (6,5) SWCNTs (bandgap: 1.3 eV) were purified by the gel chromatography [22, 27] from raw soot of eDIPS (EC1.0, Meijyo Nano Carbon Co.) and CoMoCAT (SG65, Sigma-Aldrich), respectively. Each sample was filtered through a PTFE membrane (Omnipore 0.2 μm PTFE Membrane, Merck Millipore Ltd.), and all surfactants on SWCNTs were replaced by Triton (polyoxyethylene octylphenyl ether, Wako Pure Chemical Industries, Ltd.). Then, the SWCNTs in 1% Triton solution were filtered through a polycarbonate membrane filter (Whatman Nuclepore Track-Etched membrane 25 mm 0.2 μm) to form a thin film.

**Device Preparation**

A SWCNT thin film prepared using the above procedure was transferred onto a sapphire substrate (thickness 430 ± 25 μm), which was cut to an approximate size of 2 cm × 3 cm, with predeposited gold electrodes (thickness ~100 nm). Residual polymers of a polycarbonate membrane were dissolved and washed with chloroform and acetone. The film size was approximately 3 mm (length) × 2 mm (width) × 80 ~ 100 nm (thickness). To make an electrolyte gating system, an ionic liquid (TMPA-TFSI, Kanto Chemical Co.) was dropped to cover the SWCNT film and gate electrodes, and the measurements were performed under vacuum (< 10-3 Pa).

**HHG measurements**

The laser source was a Ti:sapphire regenerative amplifier (central energy: 1.55 eV, pulse width: 35 fs, repetition rate: 1 kHz, pulse energy: 7 mJ/pulse). Part of the laser output (~1 mJ/pulse) was used to generate MIR pulses through different frequency mixing of the signal and idler outputs from an optical parametric amplifier based on an AgGaS2 crystal. The generated MIR pulses were focused onto the sample using a ZnSe lens (focal length: 60 mm). The spot size of the MIR pulse at the focal point was estimated to be 60 μm (FWHM). The HHG emissions were collected using a UV-fused silica lens (focal length: 50 mm) in transmission geometry, then spectrally resolved by a spectrometer and detected. For the detection of the $3^{rd}$-order harmonics, we used an InGaAs line detector, and the higher harmonics (>$3^{rd}$) were detected using a Si CCD camera.

**Calculation**

To simulate the effect of the Fermi level on the HHG efficiency in (6,5) SWCNTs, we calculated the temporal evolution of the density matrix and extracted the intraband current and interband polarization under MIR driving. The details of the calculations are given in the S.I. Briefly, we employed the effective Hamiltonian, which describes the (6,5) SWCNT valence and conduction bands near the K point, and then, we numerically solved the Liouville equation, which describes the electron dynamics under MIR driving, according to Ref. 26. In the simulation, we assumed a Gaussian pulse with a pulse width of 60 fs (FWHM) and a peak electric field of 1 MV/cm. The temperature $T$ and dephasing time $T_2$ were set to 300 K and 4.4 fs, respectively, and the chemical potential was systematically changed from −1 to 1 eV.

ASSOCIATED CONTENT

**Supporting Information**.

The following files are available free of charge.

The optical absorption spectra of the SWCNT samples (Figure S1), HHG spectra of Metal and (6,5) SWCNTs (Figure S2), the returnability of HHG and HHG spectra from (6,5) SWCNT as a function of gate voltages (Figure S3 and S4, respectively), relationships between HHG and dephasing time (Figure S5) and the detail of calculation (PDF)

AUTHOR INFORMATION

**Corresponding Author**

*E-mail: yanagi-kazuhiro@tmu.ac.jp.

**Author Contributions**

K.T. and K.Y. supervised the study. H.N., K.N., K.U., K.T. and K.Y. carried out the optical measurements. H.N., Y.I., Y.Y. and K.Y. contributed to the sample preparation and device fabrication. H.N., K.N., K.U., K.T. and K.Y. carried out the theoretical calculations. All authors discussed the results and contributed to the writing of the manuscript.

ACKNOWLEDGMENT

We thank Yongrui Wang, Alexey Belyanin, and Junichiro Kono for valuable discussions. K.Y. acknowledges support by JSPS KAKENHI, Grant Numbers JP17H01069, JP18H01816, JP20H02573 and JST CREST through Grant Number JPMJCR17I5, Japan. This work was supported by a Grant-in-Aid for Scientific Research (S), Grant Number JP17H06124.

ABBREVIATIONS

HHG, high-harmonic generation; MID, mid-infrared; SWCNTs, single-wall carbon nanotubes; FWHM, full width at half maximum of intensity; CCD, charge coupled device.

TOC graphics

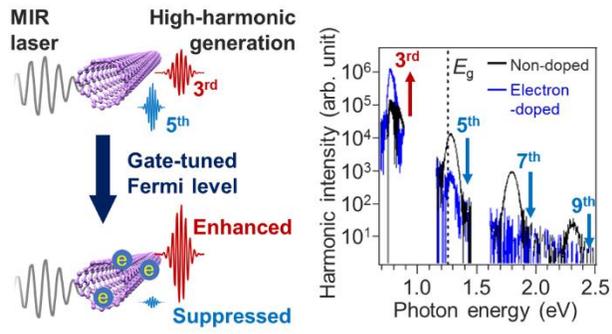

Supporting Information of

# Control of high-harmonic generation by tuning the electronic structure and carrier injection

Hiroyuki Nishidome,[1] Kohei Nagai,[2] Kento Uchida,[2] Yota Ichinose,[1] Yohei Yomogida,[1] Koichiro Tanaka,[2,3] Kazuhiro Yanagi[1*]

[1]*Department of Physics, Tokyo Metropolitan University, Hachioji, Tokyo 192-0397, Japan*

[2]*Department of Physics, Kyoto University, Sakyo-ku, Kyoto 606-8502, Japan*

[3] *Institute for Integrated Cell-Material Science (WPI-iCeMs), Kyoto University, Sakyo-ku, Kyoto 606-8501, Japan*

Corresponding Author: Kazuhiro Yanagi

*e-mail: yanagi-kazuhiro@tmu.ac.jp

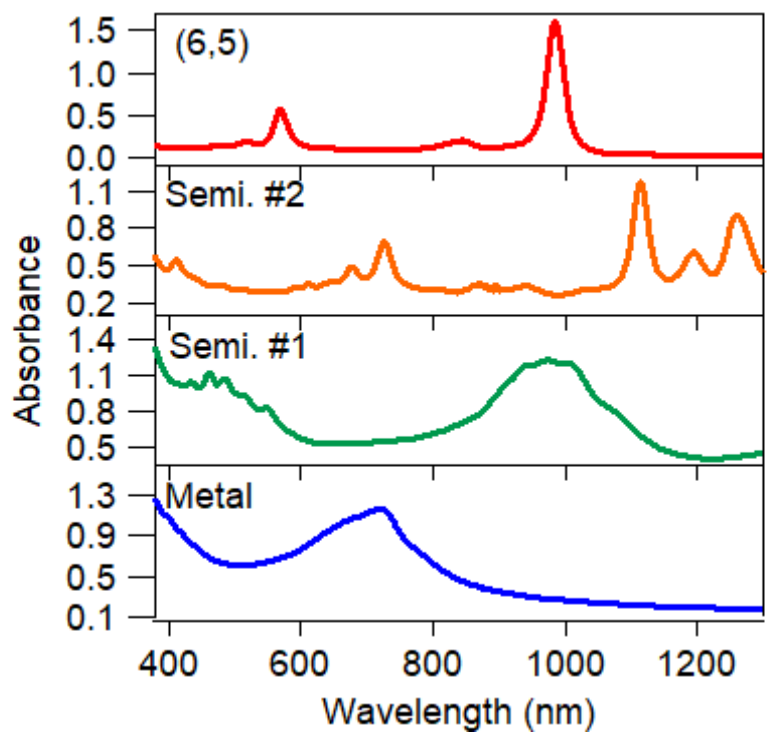

**Figure S1.** Optical absorption spectra of the Metallic (Metal) and semiconducting (Semi. #1) SWCNTs with diameter of 1.4 nm, semiconducting SWCNTs with average diameter of 1.0 nm (Semi. #2), and (6,5) SWCNTs.

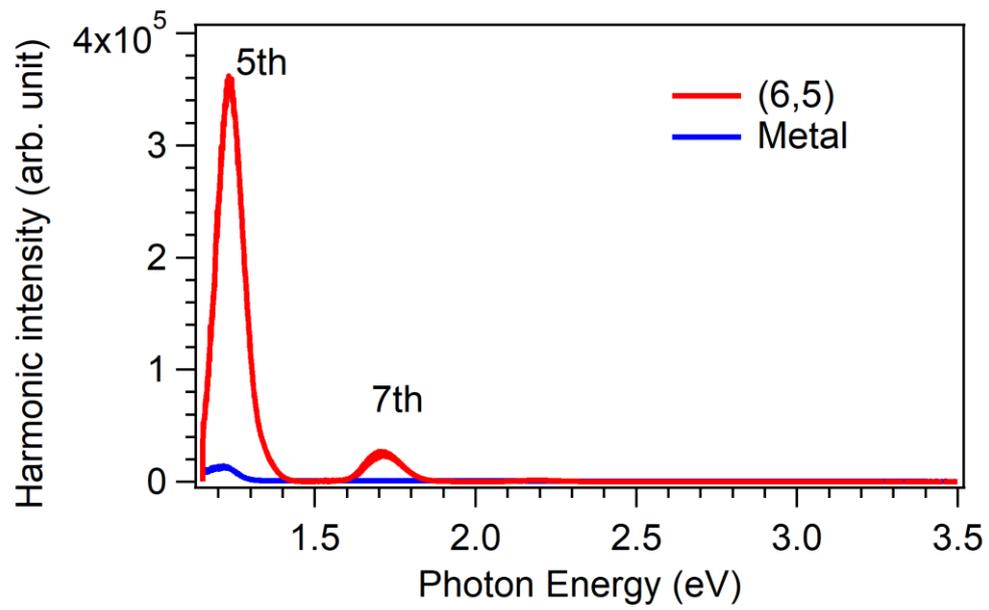

**Figure S2.** The intensity of the 5[th] order harmonics in Metal and (6,5) at the same laser power intensity. The spectra are plotted in linear scale.

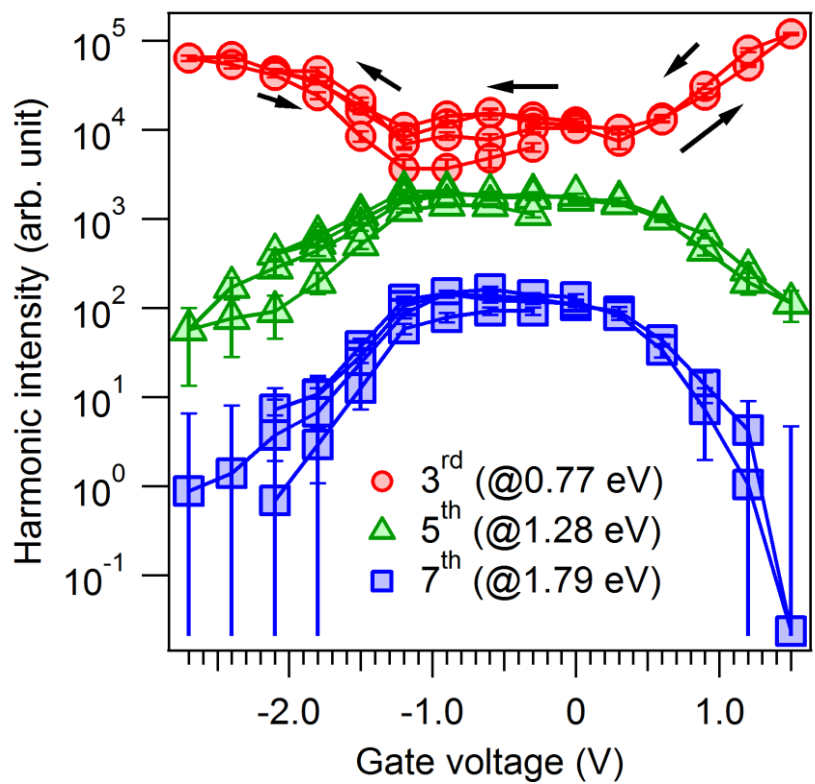

**Figure S3.** Returnability of higher order harmonics as a function of gate voltages. The arrows indicate the directions of shift of gate voltage.

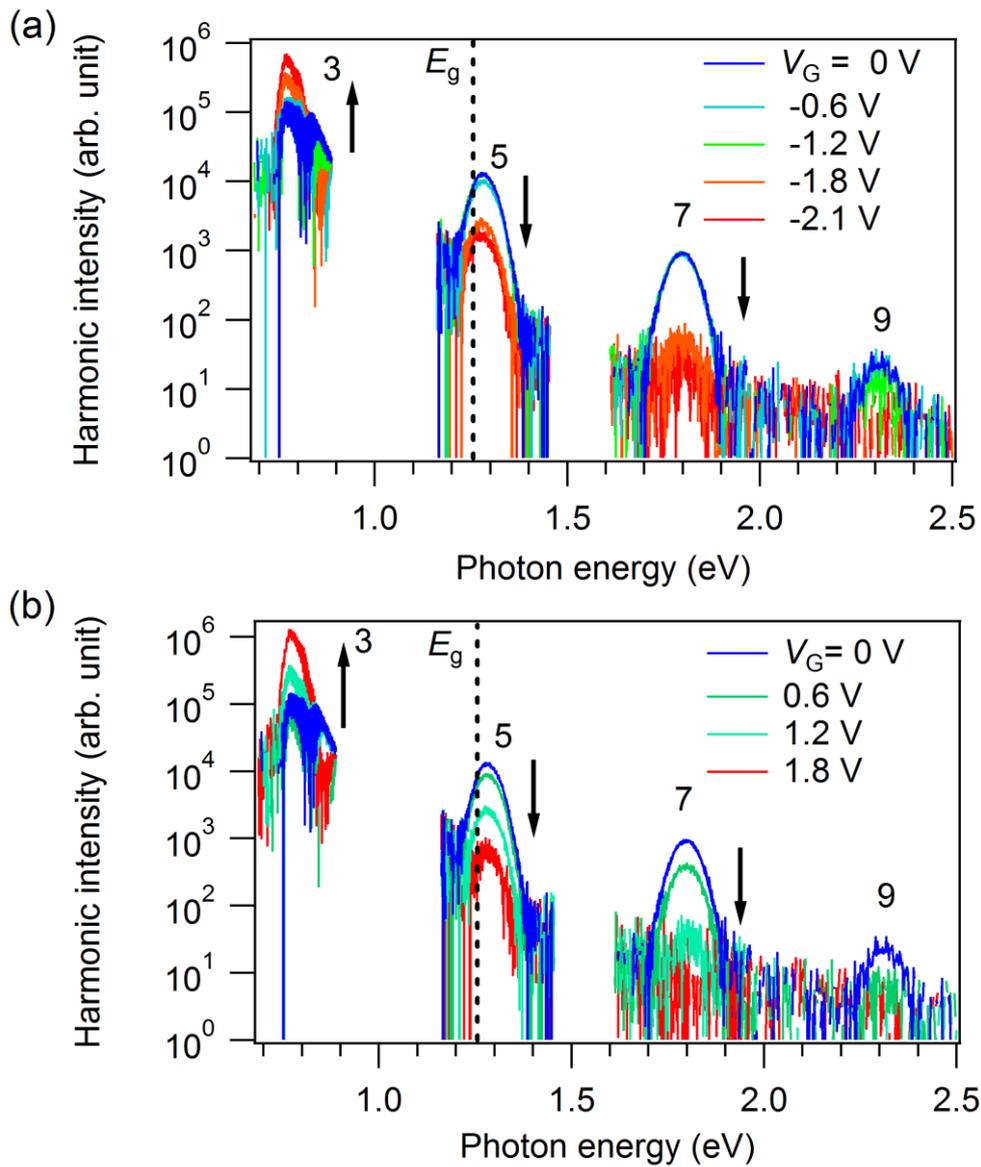

**Figure S4.** HHG spectra from (6,5) SWCNT as a function of (a) negative shift and (b) positive shift of gate voltage ($V_G$).

**The details of calculation**

To simulate the effect of Fermi levels on HHG efficiency in (6,5)-SWCNTs, we calculated the temporal evolution of the density matrix, and extracted intraband current and interband polarization under MIR driving. Here, we employ effective Hamiltonian, which describes graphene band structure near K point as follows:

$$H(k_x, k_y) = \frac{\sqrt{3}}{2}\gamma_0 a \begin{pmatrix} 0 & k_x + ik_y \\ k_x - ik_y & 0 \end{pmatrix}. \quad (S1)$$

Here, $\gamma_0$ (= 3 eV) is the hopping energy, $a$ (= 0.246 nm) is lattice constant, $\boldsymbol{k} = (k_x, k_y)$ is the crystal momentum whose origin is set to be K point. Since SWCNTs are rolled up single graphene sheet, their electronic structure near the band-minimum can be obtained from that of graphene with the periodic boundary condition along chirality vector $\boldsymbol{C}_h$.[1] Since allowed crystal momentum $\boldsymbol{k}$ must satisfy $\boldsymbol{k} \cdot \boldsymbol{C}_h = 2\pi q$ ($q$: integer), the energy band dispersion of the conduction band $E(k)$ and the transition dipole moment $d(k)$ of SWCNT near the K point are approximately given by

$$E(k_\parallel) = \frac{\sqrt{3}}{2}\gamma_0 a \sqrt{k_\parallel^2 + \Delta k_\perp^2}, \quad (S2)$$

$$d(k_\parallel) = \frac{e}{2}\frac{\Delta k_\perp}{k_\parallel^2 + \Delta k_\perp^2}, \quad (S3)$$

where $k_\parallel$ is the crystal momentum along the tube axis and $\Delta k_\perp$ is the discrete crystal momentum perpendicular to the tube axis. Here, we only consider the transition dipole moment along tube axis, and the lowest (highest) energy sub-band in the conduction (valence) band for simplicity. In this case, $\Delta k_\perp$ for (6,5)-SWCNT is given by $\Delta k_\perp = 2\pi/3a\sqrt{91}$.[1]

Based on the above parameters, we numerically solve Liouville equation, which describes electron dynamics under MIR driving, according to Ref.2:

$$\frac{\partial}{\partial t}\hat{\rho}_{k_0}(t) = -\frac{1}{i\hbar}\left[H(t), \hat{\rho}_{k_0}(t)\right], \quad (S4)$$

$$H(t) = \begin{bmatrix} -E(k_\parallel(t)) & d(k_\parallel(t))F_{MIR}(t) \\ d(k_\parallel(t))F_{MIR}(t) & E(k_\parallel(t)) \end{bmatrix}, \quad (S5)$$

$$\hat{\rho}_{k_0}(t) = \begin{bmatrix} \rho_{k_0}^{(vv)}(t) & \rho_{k_0}^{(vc)}(t) \\ \rho_{k_0}^{(cv)}(t) & \rho_{k_0}^{(cc)}(t) \end{bmatrix}, \quad (S6)$$

$$k_\parallel(t) = k_0 - \frac{e}{\hbar} A_{MIR}(t), \tag{S7}$$

where $F_{MIR}(t)$ is the temporal profile of MIR electric field, $k_0$ is the initial crystal momentum along tube axis, $A_{MIR}(t)$ is the vector potential given by the relation $F_{MIR}(t) = -\partial A_{MIR}(t)/\partial t$, and $\hat{\rho}_{k_0}(t)$ is the density matrix whose initial crystal momentum is given by $k_0$. Here, the matrix elements whose indices are $cc(vv)$ and $cv(vc)$ indicate population in the conduction (valence) band, and coherence between conduction and valence band, respectively. This describes intraband electron motion with the acceleration theorem (Eq. (S7)), and interband transition through dipole interaction (Eq. (S5)). We assume that the initial conditions of density matrix obey the Fermi-Dirac distribution as follows:

$$\rho_{k_0}^{(vv)}(0) = \frac{1}{\exp\left\{\frac{(-E(k_0)-\mu)}{k_B T}+1\right\}}, \tag{S8}$$

$$\rho_{k_0}^{(cc)}(0) = \frac{1}{\exp\left\{\frac{(E(k_0)-\mu)}{k_B T}+1\right\}}, \tag{S9}$$

$$\rho_{k_0}^{(vc)}(0) = \rho_{k_0}^{(cv)}(0) = 0. \tag{S10}$$

Here, $T$ is the temperature, and $\mu$ is the chemical potential.

The obtained temporal evolution of density matrix gives intraband current and interband polarization described by

$$J(t) = \frac{e}{\hbar} \int dk_0 \left\{ \left.\frac{\partial E(k_\parallel)}{\partial k_\parallel}\right|_{k_\parallel=k_\parallel(t)} \rho_{k_0}^{(cc)}(t) - \left.\frac{\partial E(k_\parallel)}{\partial k_\parallel}\right|_{k_\parallel=k_\parallel(t)} \rho_{k_0}^{(vv)}(t) \right\}, \tag{S11}$$

$$P(t) = \int dk_0 \, d\left(k_\parallel(t)\right) \left\{ \rho_{k_0}^{(vc)}(t) + \rho_{k_0}^{(cv)}(t) \right\}, \tag{S12}$$

where the integral is taken from $-2\pi/\sqrt{3}a$ to $2\pi/\sqrt{3}a$ in our calculation. Finally, we obtain HHG spectrum from the result of Eqs. (S11) and (S12) given by

$$I_{HHG}(\omega) = |J(\omega) + i\omega P(\omega)|^2. \tag{S13}$$

Note that we empirically introduced the damping of coherence given by $-\rho_{k_0}^{(ij)}(t)/T_2$ in Eq. (S4), where ($ij$) is ($vc$) or ($cv$), and $T_2$ is the dephasing time of interband polarization.

In the simulation shown in the main text, we assume Gaussian pulse with the pulse width of 60 fs (full width of half maximum of intensity), and the peak electric field of 1 MV/cm. Temperature $T$ and dephasing time $T_2$ is set to be 300 K and 4.4 fs respectively, and chemical potential $\mu$ is systematically changed from $-1$ to 1 eV. The dephasing time influences the intensities of 3rd, 5th and 7th order harmonics, as shown in Figure S5. The experimentally observed intensity order is 3rd > 5th > 7th. Thus we selected 4.4 fs for $T_2$ so that the calculated intensity order agrees with the experimental observation.

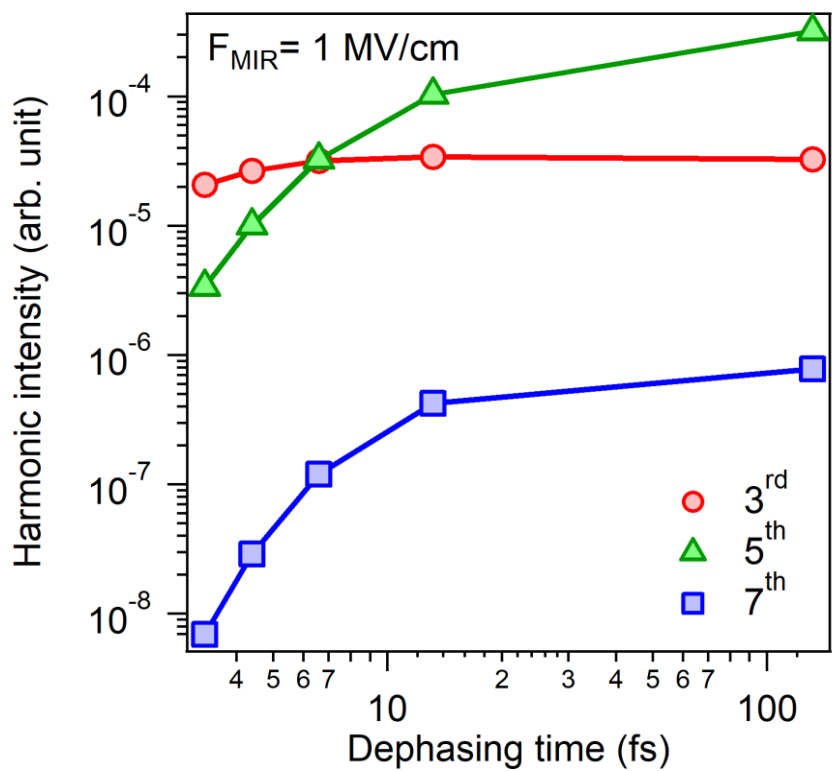

**Figure S5.** Relationships among 3rd, 5th, 7th harmonic intensities and dephasing time.